\documentclass[twocolumn,aps,pra,superscriptaddress]{revtex4}
\usepackage[latin9]{inputenc}
\usepackage{amsmath}
\usepackage{amssymb}
\usepackage{graphicx}

\makeatletter

\@ifundefined{textcolor}{}
{%
 \definecolor{BLACK}{gray}{0}
 \definecolor{WHITE}{gray}{1}
 \definecolor{RED}{rgb}{1,0,0}
 \definecolor{GREEN}{rgb}{0,1,0}
 \definecolor{BLUE}{rgb}{0,0,1}
 \definecolor{CYAN}{cmyk}{1,0,0,0}
 \definecolor{MAGENTA}{cmyk}{0,1,0,0}
 \definecolor{YELLOW}{cmyk}{0,0,1,0}
}

\usepackage{color}
\usepackage{braket}

\usepackage{subfigure}\usepackage{epsfig}\usepackage{amsfonts}\usepackage{mathrsfs}\usepackage[toc,page,title,titletoc,header]{appendix}

\makeatother

\begin{document}


\title{Atom-dimer scattering in heteronuclear mixture with finite intra-species
scattering length}



\author{Chao Gao}
\email[]{gaochao42@gmail.com}

\affiliation{Department of Physics, Zhejiang Normal University, Jinhua, 321004,
China}

\affiliation{Department of Physics, Renmin University of China, Beijing, 100872,
China}


\author{Peng Zhang}

\email[]{pengzhang@ruc.edu.cn}

\affiliation{Department of Physics, Renmin University of China, Beijing, 100872,
China}

\affiliation{Beijing Computational Science Research Center, Beijing, 100084, China}

\affiliation{Beijing Key Laboratory of Opto-electronic Functional Materials \&
Micro-nano Devices (Renmin University of China)}




\date{today}

\begin{abstract}
We study the three-body problem of two ultracold identical bosonic atoms (denoted by $B$) and one extra atom (denoted by $X$), where the scattering length $a_{BX}$ between each bosonic atom and atom $X$ is resonantly large and positive. We calculate the scattering length $a_{{\rm ad}}$ between one bosonic atom and the shallow dimer formed by the other bosonic atom and atom $X$, and investigate the effect induced by the interaction between the two bosonic atoms. We find that even if this interaction is weak (i.e., the corresponding scattering length $a_{BB}$ is of the same order of the van der Waals length $r_{{\rm vdW}}$ or even smaller), it can still induce a significant effect for the atom--dimer scattering length $a_{{\rm ad}}$. Explicitly, an atom--dimer scattering resonance can always occur when the value of $a_{BB}$ varies in the region with $|a_{BB}|\lesssim r_{{\rm vdW}}$. As a result, both the sign and the absolute value of $a_{{\rm ad}}$, as well as the behavior of the $a_{{\rm ad}}$-$a_{BX}$ function, depends sensitively on the exact value of $a_{BB}$. Our results show that, for a good quantitative theory, the intra-species interaction is required to be taken into account for this heteronuclear system, even if this interaction is weak. 
\end{abstract}
\maketitle

\section{Introduction}

The scattering problem between a shallow dimer and a single atom is important for both few-body and many-body physics of ultracold atom gases. For few-body physics, the atom--dimer scattering is closely related to the Efimov effect \cite{Efimov1970,Efimov1971,Efremov2009,HHP2010}, ultracold chemistry processes \cite{bozhao}, and three-body loss \cite{shizhong}. For many-body problems, the atom--dimer scattering length determines the interaction energy in the mixture of ultracold atoms and shallow dimers \cite{Cui2014,Zhang2014}. Therefore, the atom--dimer scattering problem has been investigated recently for various types of ultracold gases. In Table I, we list all the references we have found for the calculation of the scattering length between an atom and a shallow dimer in ultracold gases.
\begin{center}
\begin{table}[!hbp]
\begin{center}
\begin{tabular}{c l}
\hline
\hline
Type &\ \  Reference\\
\hline
$B-BB$  & \ \ V. Efimov, 1979 \cite{Efimov1979} \\
 & \ \ P. F. Bedaque \textit{et al.}, 1999 \cite{Bedaque1999} \\
  & \ \ D. S. Petrov, 2004 \cite{Petrov2004BBB} \\
\hline 
$F-FX$  & \ \ G. V. Skorniakov \textit{et al.}, 1957 \cite{STM1957} \\
 & \ \ D. S. Petrov, 2003 \cite{Petrov2003} \\
 & \ \ D. S. Petrov \textit{et al}, 2005 \cite{Petrov2005} \\
 & \ \ J. Levinsen \textit{et al}, 2009 \cite{Levinsen2009}; 2011
\cite{Levinsen2011}\\
 & \ \ M. Iskin, 2010 \cite{Iskin2010} \\
 & \ \ F. Alzetto \textit{et al}, 2010 \cite{Alzetto2010}; 2012 \cite{Alzetto2012} \\
 & \ \ S. Bour \textit{et al}, 2012 \cite{Bour2012} \\
\hline 
$B-BX$  & \ \ M. A. Efremov \textit{et al}, 2009 \cite{Efremov2009}\\
 & \ \ K. Helfrich \textit{et al}, 2010 \cite{HHP2010} \\
 & \ \ B. Acharya \textit{et al}, 2016 \cite{AJP2016} \\
\hline 
$X-YZ$  & \ \ X. Cui, 2014 \cite{Cui2014} \\
 & \ \ R. Zhang \textit{et al}, 2014 \cite{Zhang2014} \\
\hline 
\end{tabular}
\end{center}
\caption{References on the calculation of scattering length between an ultracold atom and a shallow dimer. Here $B$ and $F$ denote identical bosonic and fermonic atoms, respectively, and $X$, $Y$, and $Z$ denote distinguishable atoms. For instance, ``$B-BB$\char`\"{} is the scattering between one bosonic atom and a dimer formed by two identical bosonic atoms (i.e., all the three atoms are identical); ``$F-FX$\char`\"{} is the scattering between one fermonic atom and a dimer formed by one identical fermonic atom and a distinguishable atom.}
\end{table}
\par\end{center}

\begin{figure}
\includegraphics[width=4.5cm]{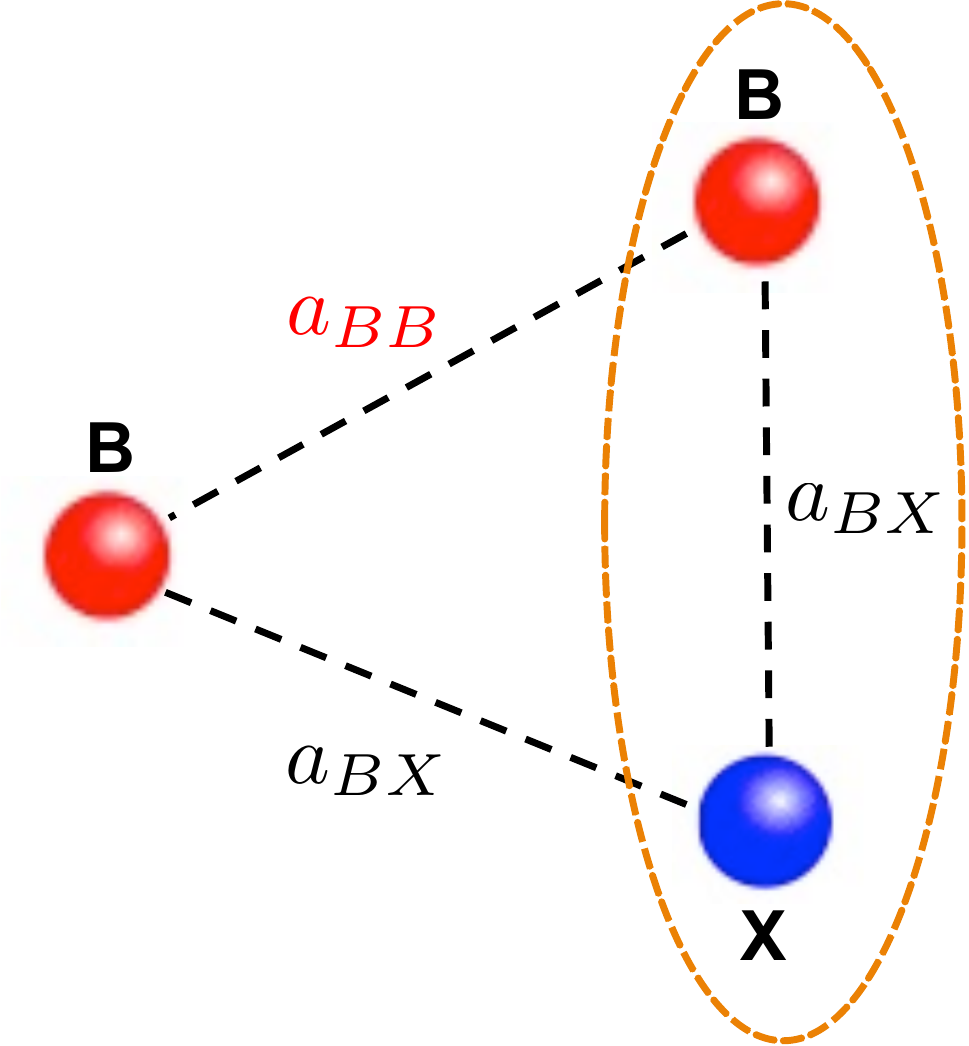} 
\caption{(color online) Schematic of the three-body system. We consider two bosonic atoms of type $B$ and one extra atom $X$, and assume the scattering length $a_{BX}$ is positive and much larger than the van der Waals length. We calculate the scattering length $a_{{\rm ad}}$ between one bosonic atom and the shallow dimer formed by atom $X$ and the other bosonic atom, and investigate the effect induced by the scattering length $a_{BB}$ between the bosonic atoms.}
\label{scheme} 
\end{figure}

In our study, we calculated the scattering length $a_{{\rm ad}}$ between a bosonic atom of type $B$ and a shallow dimer formed by an identical bosonic atom, also of type $B$, and one distinguishable atom (denoted by $X$); see Fig.~\ref{scheme}. This system can be experimentally realized in the ultracold gases of the $^{23}$Na$-$$^{40}$K mixture \cite{bozhao}, $^{85}$Rb$-$$^{7}$Li mixture \cite{Rb85Li7}, $^{87}$Rb$-$$^{41}$K mixture \cite{RbK41}, $^{87}$Rb$-$$^{40}$K mixture \cite{RbK40a,RbK40,RbK40b}, $^{87}$Rb$-$$^{87}$Sr/$^{88}$Sr mixture \cite{RbSr}, and $^{133}$Cs$-$$^{6}$Li mixture \cite{LiCs,LiCsb,LiCsc,LiCsd,LiCse}. For this system, the intensities of the interaction between the two bosonic atoms and the interaction between the bosonic atom and the atom $X$ are described by the corresponding scattering lengths $a_{BB}$ and $a_{BX}$, respectively. Furthermore, $a_{BX}$ is positive and much larger than the range of the interaction, i.e., the van der Waals length $r_{{\rm vdW}}$. For the case with $a_{BB}=0$, M. A. Efremov {\it et. al.},\cite{Efremov2009} and K. Helfrich {\it et. al.}, \cite{HHP2010} have calculated the atom--dimer scattering length using the Born--Oppenheimer approximation and exact numerical calculation, respectively. Nevertheless, in most realistic systems, $a_{BB}$ is non-zero. In 2016, Acharya, Ji, and Platter treated the boson--boson interaction as a perturbation and derived a formal solution for the atom--dimer scattering amplitude \cite{AJP2016}. However, to the best of our knowledge, a quantitative numerical calculation for the atom--dimer scattering length $a_{{\rm ad}}$ for finite $a_{BB}$ is still absent.

In this work, we give such a calculation via the Skorniakov--Ter--Martirosian (STM) equation \cite{STM1957} approach, and investigate the effect of finite $a_{BB}$ on $a_{{\rm ad}}$. We focus on systems with weak boson--boson interaction, i.e., $|a_{BB}|$ is of the same order of $r_{{\rm vdW}}$ or even smaller. This appears also the scenario in most of the current experiments. It has been shown that this boson--boson interaction modifies the spectrum of three-body bound state and Bose polaron for the systems with $a_{BX}<0$ \cite{LiCsc,LiCsb,LiCse}. In this work, we find that, although the boson--boson interaction is weak, it can still induce a significant effect for atom--dimer scattering. Explicitly, we find that $a_{{\rm ad}}$ changes resonantly with $a_{BB}$, even in the small region $|a_{BB}|\lesssim r_{\rm vdW}$. Furthermore, this atom--dimer resonance, which is induced by the variation of $a_{BB}$, can always occur for various values of $a_{BX}$, and therefore is a universal feature of this system. This resonance effect not only influences the value of $a_{{\rm ad}}$, but also modifies the behavior of the $a_{{\rm ad}}$-$a_{BX}$ function. In particular, when $a_{BB}$ is varied in the region $|a_{BB}|\lesssim r_{\rm vdW}$, the position of the resonance points of the $a_{{\rm ad}}$-$a_{BX}$ function can be changed by several orders of magnitudes. Our results show that even if the boson--boson interaction is weak in our system, it still needs to be taken into account in theoretical treatments.

The remainder of this paper is organized as follows. In Sec.~II, we describe our theoretical model and the STM equation. In Sec.~III, we present our results regarding $a_{{\rm ad}}$ and analyze the results induced by the finite value of $a_{BB}$. A summary and discussion are given in Sec.~IV. In the appendix we present details of the derivation of the STM equation.



\section{STM equation}

We consider the three-body system with two identical bosonic atoms and one distinguishable atom $X$, with the inter-species scattering length $a_{BX}$ being large and positive (Fig.~\ref{scheme}). In consequence, atom $X$ and one of the bosonic atoms may form a shallow dimer with energy 
\begin{eqnarray}
E_{b}=-\frac{\hbar^{2}}{2\mu_{BX}a_{BX}^{2}},
\end{eqnarray}
where $\mu_{BX}=m_{B}m_{X}/(m_{B}+m_{X})$ is the reduced mass of one bosonic atom and atom $X$, with $m_{B}$ and $m_{X}$ being their respective masses. We calculate the scattering length $a_{{\rm ad}}$ between the shallow dimer and the other bosonic atom via the STM equation approach \cite{STM1957}. In our calculation, we describe the boson--boson interaction and the boson--$X$ interaction with zero-range Huang--Yang pseudo-potentials $V_{BB}$ and $V_{BX}$, respectively. They are given by 
\begin{eqnarray}
V_{BB} & = & \frac{2\pi\hbar^{2}a_{BB}}{\mu_{BB}}\delta({\bf r}_{BB})\frac{\partial}{\partial r_{BB}}(r_{BB}\cdot);\\
V_{BX} & = & \frac{2\pi\hbar^{2}a_{BX}}{\mu_{BX}}\delta({\bf r}_{BX})\frac{\partial}{\partial r_{BX}}(r_{BX}\cdot),
\end{eqnarray}
with $\mu_{j}$ and ${\bf r}_{j}$ ($j=BB,BX$) being the corresponding two-body reduced mass and relative position. Using this model, we can derive by a direct calculation (Appendix A) the STM equation for our system ($\hbar=m_{B}=1$):
\begin{widetext}
\begin{eqnarray}
 &  & \frac{2^{1/2}\mu_{BX}^{1/2}A(K,\varepsilon)}
{\left(\frac{K^{2}}{2m_{{\rm ad}}}+|E_{b}|-i\varepsilon\right){}^{1/2}+|E_{b}|^{1/2}}
-\frac{m_{X}/m_{B}+1}{2\pi K}\int_{0}^{\Lambda e^{i\zeta}}dK'
\frac{K'A\left(K',\varepsilon\right)}{\frac{K'^{2}}{2m_{\text{ad}}}-i\varepsilon}
\ln\left(\frac{\frac{K^{\prime2}+K^{2}}{2\mu_{BX}}+\frac{K'K}{M}+|E_{b}|-i\varepsilon}
{\frac{K^{\prime2}+K^{2}}{2\mu_{BX}}-\frac{K'K}{M}+|E_{b}|-i\varepsilon}\right)\nonumber \\
 &  & -\frac{2^{3/2}\pi^{1/2}m_{\text{ad}}\sqrt{a_{BX}}}{K}
\int_{0}^{\Lambda e^{i\zeta}}dK'K'\eta\left(K',\varepsilon\right)
\ln\left(\frac{\frac{K^{\prime2}}{2\mu_{BX}}+K^{2}+K'K+|E_{b}|-i\varepsilon}
{\frac{K^{\prime2}}{2\mu_{BX}}+K^{2}-K'K+|E_{b}|-i\varepsilon}\right)\nonumber \\
 &  & =-\frac{m_{\text{ad}}}{\mu_{BX}\left(\frac{K^{2}}{2\mu_{BX}}+|E_{b}|-i\varepsilon\right)}+i\varepsilon\sqrt{\frac{\pi}{2a_{BX}}}\frac{1}{[-|E_{b}|\mu_{BX}+i\varepsilon\mu_{BX}-|{\bf K}|^{2}/2](|{\bf K}|^{2}+|E_{b}|)};\label{stm1}
\end{eqnarray}
\begin{equation}
\begin{aligned} & -\frac{1}{2\pi m_{\text{ad}}K}\int_{0}^{\Lambda e^{i\zeta}}dK'
\frac{K'A\left(K',\varepsilon\right)}{\frac{K'^{2}}{2m_{\text{ad}}}-i\varepsilon}
\ln\left(\frac{K^{\prime2}+K'K+\frac{K^{2}}{2\mu_{BX}}+|E_{b}|-i\varepsilon}
{K^{\prime2}-K'K+\frac{K^{2}}{2\mu_{BX}}+|E_{b}|-i\varepsilon}\right)\\
 & +2^{1/2}\pi^{3/2}\mu_{BX}\sqrt{a_{BX}}\left[
\left(\frac{K^{2}}{4\mu_{BX}m_{\text{ad}}}+|E_{b}|-i\varepsilon\right){}^{1/2}-a_{BB}^{-1}
\right]\eta(K,\varepsilon)=-\frac{1}{\frac{K^{2}}{2\mu_{BX}}+|E_{b}|-i\varepsilon},
\end{aligned}
\label{stm2}
\end{equation}
\end{widetext}
with $m_{\text{ad}}=(m_{X}/m_{B}+1)/(m_{X}/m_{B}+2)$ being the value of the atom--dimer reduced mass in our natural unit. As shown in Appendix A, the atom--dimer scattering length $a_{{\rm ad}}$ is related to the solution $\{A(K,\epsilon),\eta(K,\epsilon)\}$ of the STM equation via the relation 
\begin{eqnarray}
a_{{\rm ad}}=\lim_{\epsilon\to0^{+}}A(K=0,\epsilon).\label{sl}
\end{eqnarray}

In the STM equations (\ref{stm1}) and (\ref{stm2}), the momentum cutoff $\Lambda e^{i\zeta}$ is introduced to regularize the integration. This parameter describes the physics when all three atoms are close to each other. The exact value of $\Lambda$ and $\zeta$ is determined by the short-range detail of the two-atom interaction. It is known that $1/\Lambda$ is usually of the order of the van der Waals length $r_{{\rm vdW}}$ whereas the phase angle $\zeta$, which describes the three-body recombination process, is usually very small. Thus, for our system, we have $a_{BX}\gg1/\Lambda$. In this study, we consider systems with $a_{BX}>30/\Lambda$. Furthermore, we also assume that $a_{BB}$ is comparable or smaller than $r_{{\rm vdW}}$, which implies $|a_{BB}|\Lambda\lesssim1$. Therefore, we consider the instance with $a_{BB}\in\left[-3/\Lambda,+3/\Lambda\right]$.

In our problem, the atom--dimer scattering length $a_{{\rm ad}}$ has a small imaginary part when $\zeta\neq0$ or $a_{BB}>0$. In the former case, ${\rm Im}[a_{{\rm ad}}]$ is induced by three-body recombination processes. In the latter case, for our model, the two bosonic atoms can form a deep bound state with energy $-1/a_{BB}^{2}$, and the atoms may decay to this deep bound state after atom--dimer scattering. This decay process can also induce a non-zero value for ${\rm Im}[a_{{\rm ad}}]$. In the following, we focus on the behavior of the real part of $a_{{\rm ad}}$, which describes the intensity of the atom--dimer interaction. As shown below, our results are insensitive to the value of $\zeta$.

\section{$a_{BB}$-induced effect for $a_{{\rm ad}}$}

We numerically solve the STM equations (\ref{stm1}, \ref{stm2}) and derive the atom--dimer scattering length $a_{{\rm ad}}$ via the relation (\ref{sl}). In this section, we show our results and investigate the effect of the finite $a_{BB}$ on $a_{{\rm ad}}$. Here we consider the systems with mass ratio $m_{B}:m_{X}=23:40$, $1:1$, and $87:40$ (corresponding to the $^{23}$Na$-^{40}$K mixture, $^{87}$Rb$-^{87}$Sr mixture, and $^{87}$Rb$-^{40}$K mixture, respectively). In the following, we first investigate the dependence of $a_{{\rm ad}}$ on $a_{BB}$, in instances with fixed $a_{BX}$, and then study the influence of $a_{BB}$ on the behavior of the $a_{{\rm ad}}$-$a_{BX}$ function.

\subsection{Relationship between $a_{{\rm ad}}$ and $a_{BB}$}

As shown above, in our system with $|a_{BB}|\ll a_{BX}$, the boson--boson interaction $V_{BB}$ is much weaker than the inter-species interaction $V_{BX}$. Nevertheless, our results show that the variation of $a_{BB}$ significantly modifies the atom--dimer scattering length $a_{{\rm ad}}$. 

\begin{figure*}[t]
\includegraphics[width=6.5cm]{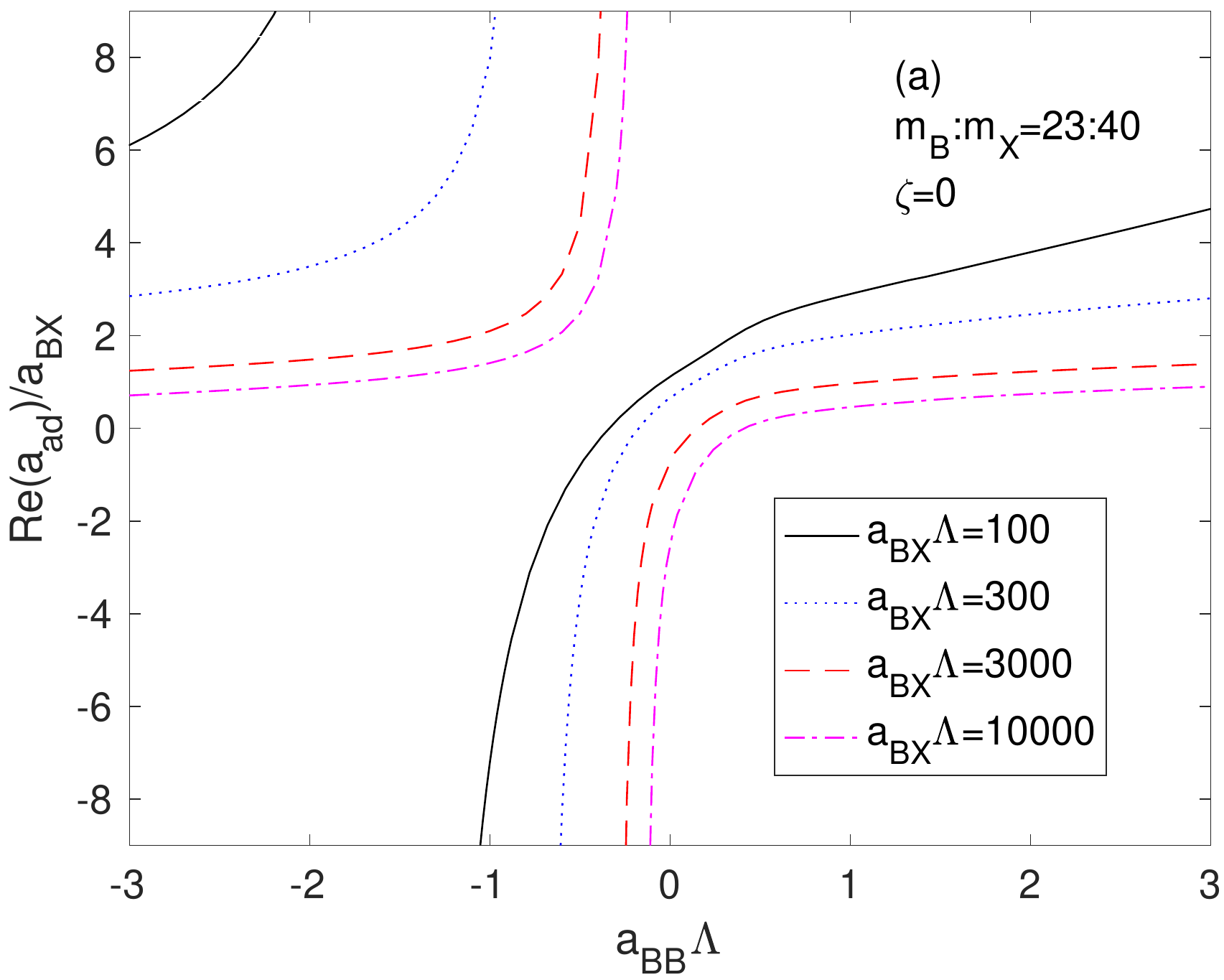} \includegraphics[width=6.5cm]{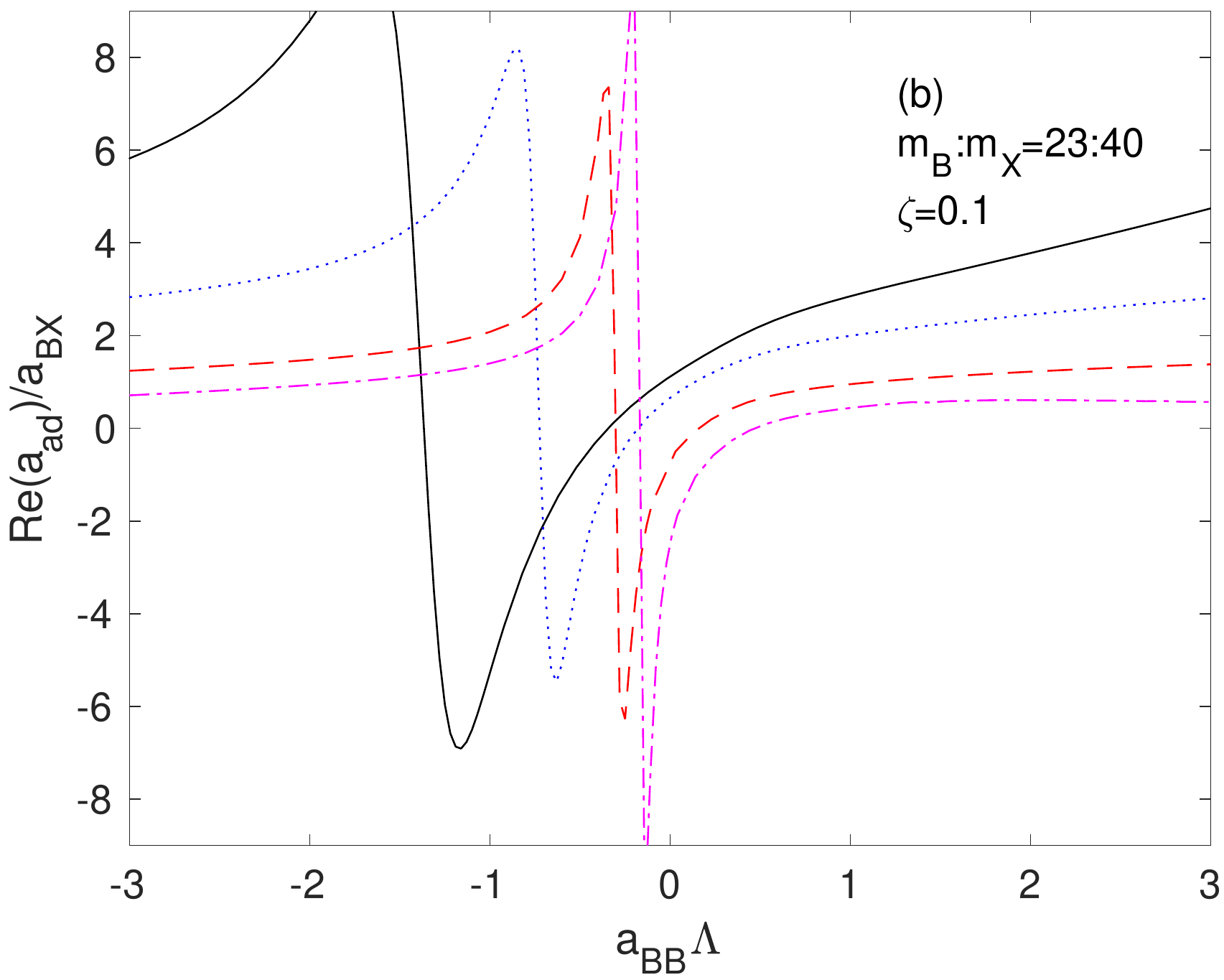}
\includegraphics[width=6.5cm]{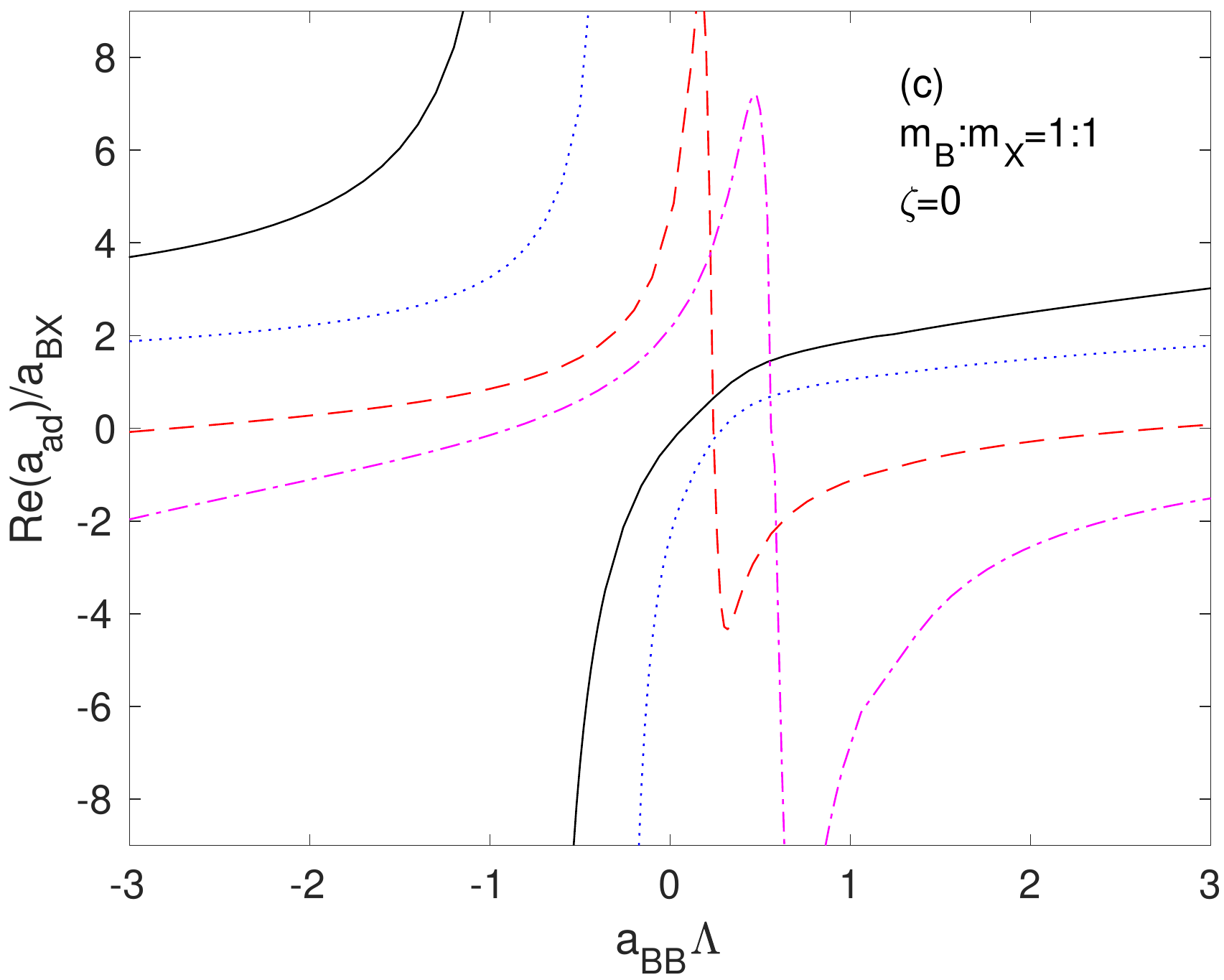} \includegraphics[width=6.5cm]{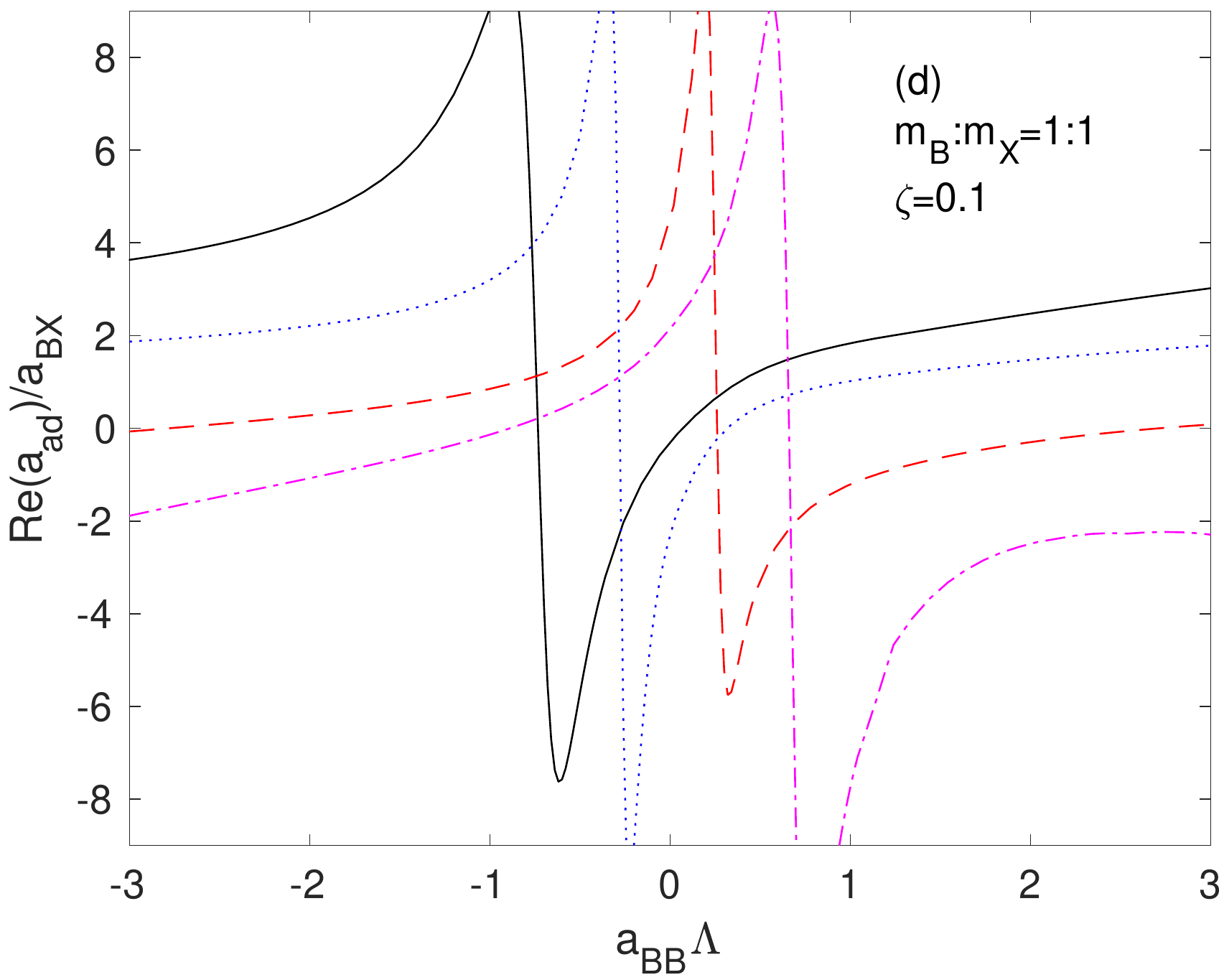}
\includegraphics[width=6.5cm]{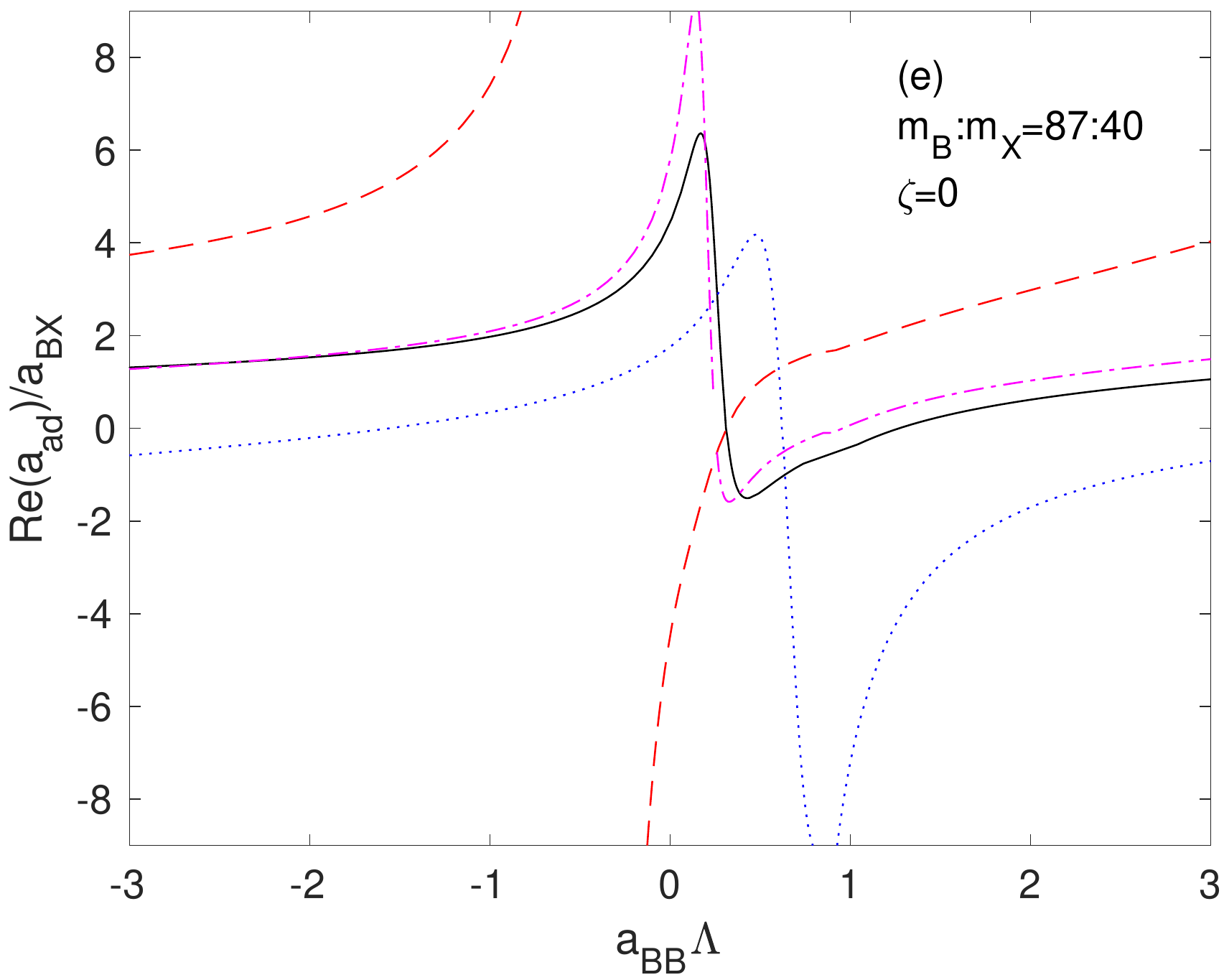} \includegraphics[width=6.5cm]{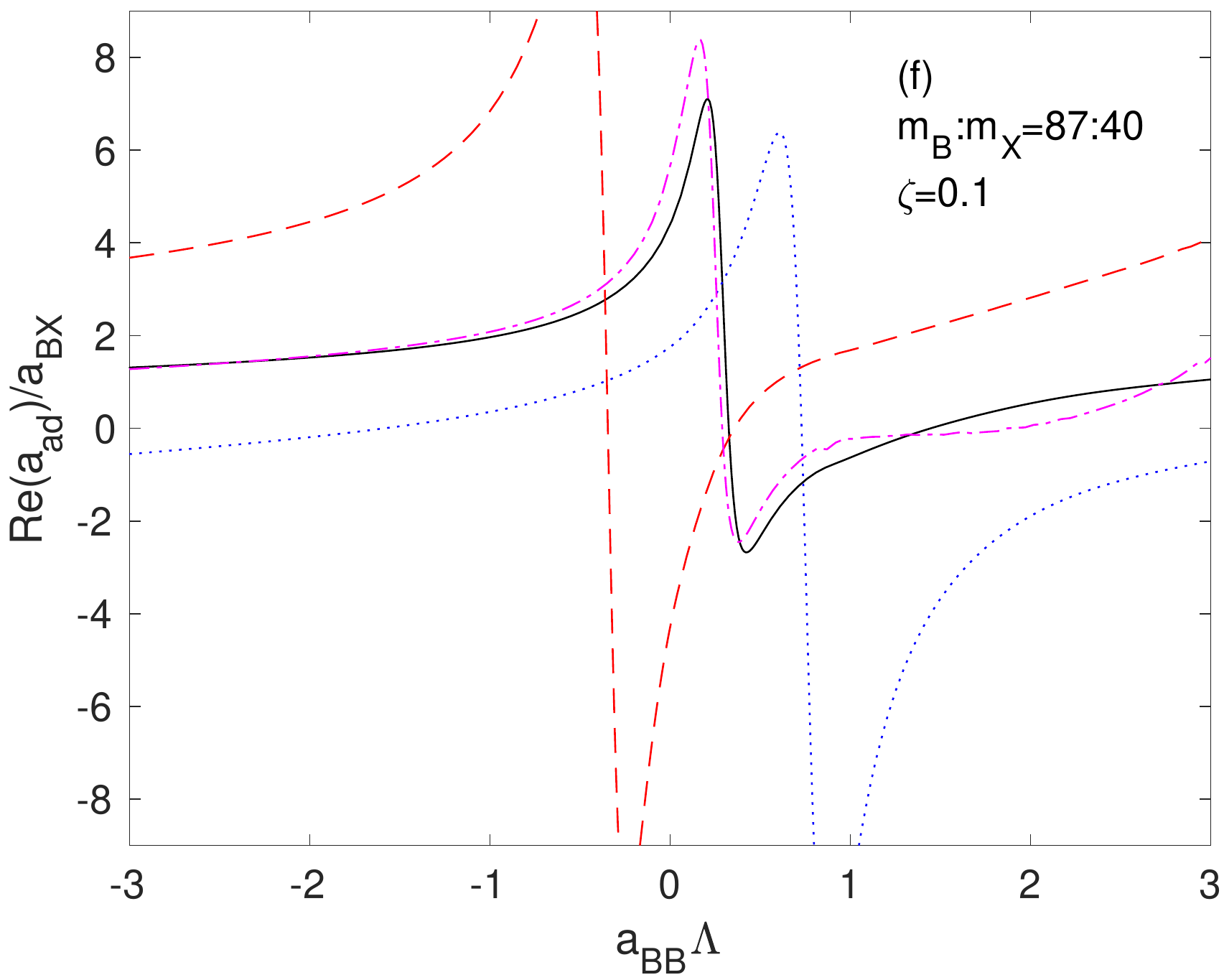}
\caption{(color online) ${\rm Re}[a_{{\rm ad}}]/a_{BX}$ as a function of $a_{BB}$ for $m_{B}:m_{X}=23:40$ (a, b), $1:1$ (c, d), and $87:40$ (e, f), with $\zeta=0$ and $\zeta=0.1$. For each case, we show the results for $a_{BX}=100/\Lambda$ (black solid line), $300/\Lambda$ (blue dotted line), $3000/\Lambda$ (red dashed line) and $10000/\Lambda$ (magenta dashed-dotted line).}
\label{aadabb} 
\end{figure*}

In Fig.~\ref{aadabb}, we plot ${\rm Re}[a_{{\rm ad}}]$ as a function of $a_{BB}$, for cases with different mass ratio $m_B/m_X$ and different inter-species scattering length $a_{BX}$. For each case, we illustrate the results for $\zeta=0$ and $\zeta=0.1$, with $\zeta$ being the phase of the momentum cut-off, as defined above. The variation of $a_{BB}$ in the small region $a_{BB}\in\left[-3/\Lambda,+3/\Lambda\right]$ clearly always induces a \textit{resonance} of the atom--dimer scattering. In addition, when $\zeta=0$, if the resonance appears for $a_{BB}<0$, then ${\rm Re}[a_{{\rm ad}}]$ diverges at the resonance point. Nevertheless, ${\rm Re}[a_{{\rm ad}}]$ no longer diverges for all cases with $a_{BB}>0$ or $\zeta\neq 0$, even at the resonance point. This is due to the presence of inelastic processes, i.e., three-body recombinations or inelastic scatterings to the deep bound state formed by the two bosons. For all these cases, we define the resonance of ${\rm Re}[a_{{\rm ad}}]$ as the centroid of the positions of the positive and negative maximum values of ${\rm Re}[a_{{\rm ad}}]$. As shown in Fig.~\ref{aadabb}, the resonance-point positions, as well as other behaviors of ${\rm Re}[a_{{\rm ad}}]$, are quite similar for $\zeta=0$ and $\zeta=0.1$.

This result is further confirmed by Fig.~\ref{resonance2}, where we plot the resonance position of the ${\rm Re}[a_{{\rm ad}}]$--$a_{BB}$ curve (i.e., the value of $a_{BB}$ for which resonance can occur) for various values of $a_{BX}$. This figure shows that the atom--dimer scattering resonances always occurs when $a_{BB}$ ranges within the region $|a_{BB}|\lesssim1/\Lambda$. Therefore, the resonances induced by the variation of the small boson--boson scattering length is a universal feature of the atom--dimer scattering in this system.

Because of this resonance effect, the atom--dimer scattering length $a_{{\rm ad}}$ depends sensitively on the value of $a_{BB}$. A slight variation of $a_{BB}$ may significantly change the value of $a_{{\rm ad}}$. Thus, although the boson--boson interaction is weak, it cannot be simply ignored in a quantitative calculation.

We may understand the ``$a_{BB}$-induced\char`\"{} atom--dimer resonance with the following physical picture. In our system, there is a series of three-body bound states, i.e., the Efimov bound states. An atom--dimer resonance appears when the energy $E_{{\rm bound}}^{({\rm 3-body)}}$
of the three-body bound state in consonance with the energy of the dimer, i.e.,
\begin{eqnarray}
E_{{\rm bound}}^{{\rm (3-body)}}=E_{b}.\label{adresonance}
\end{eqnarray}
Furthermore, the energy of the three-body bound states depends on the exact value of $a_{BB}$. Therefore, the results displayed in Figs.~\ref{aadabb} and \ref{resonance2} implies that the variation of $a_{BB}$ in the small region with $|a_{BB}|\lesssim1/\Lambda$ always shifts the energy of one three-body bound state to ensure the condition (\ref{adresonance}) is satisfied.

\subsection{Relationship between $a_{{\rm ad}}$ and $a_{BX}$}

Now we consider the relationship between the atom--dimer scattering length $a_{{\rm ad}}$ and the scattering length $a_{BX}$ between the bosonic atom and atom $X$. For $a_{BB}=0$, K. Helfrich {\it et. al.} \cite{HHP2010} showed that the ratio between $a_{{\rm ad}}$ and $a_{BX}$ satisfies Efimov's radial law \cite{Efimov1979}
\begin{eqnarray}
\frac{a_{{\rm ad}}}{a_{BX}}=C_{1}+C_{2}\cot\left[s_{0}\ln\left(a_{BX}\Lambda\right)+\phi\right].
\label{ERLeq}
\end{eqnarray}
Here $C_{1}$, $C_{2}$, $s_{0}$, and $\phi$ are functions of the mass ratio $m_{B}/m_{X}$, and are independent of the values of $a_{BX}$ and $\Lambda$. As a result, the atom--dimer scattering resonances appear when
\begin{eqnarray}
a_{BX}=\frac{e^{\frac{(n\pi-\phi)}{s_{0}}}}{\Lambda}\ \ (n=0,\pm1,\pm2,...).
\end{eqnarray}

\begin{figure}[t]
\includegraphics[width=7cm]{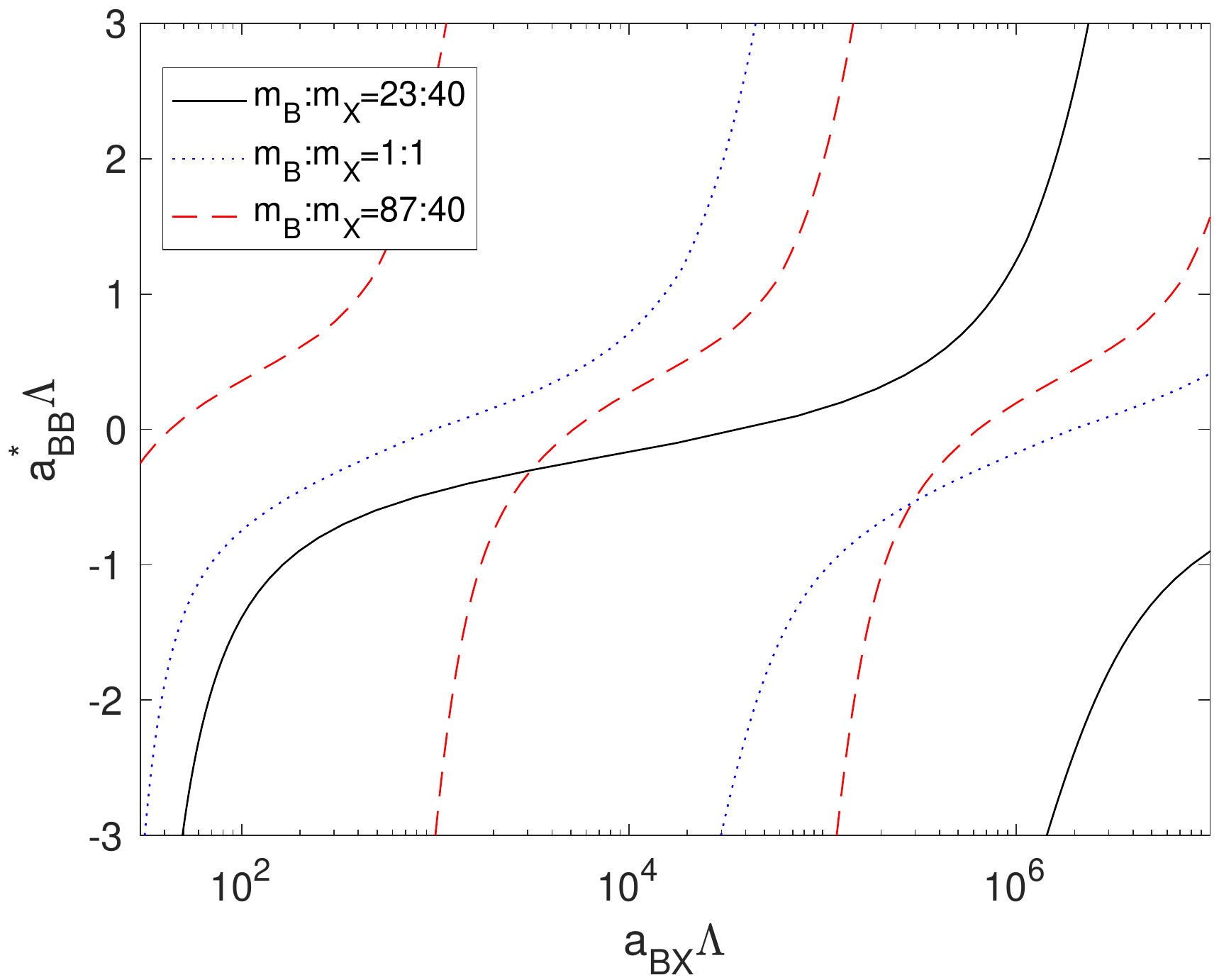}
\caption{(color online) Resonance position $a_{BB}^{\ast}$ of the ${\rm Re}[a_{{\rm ad}}]$-$a_{BB}$ function (i.e., $a_{BB}^{\ast}$ is the value of $a_{BB}$ for which a resonance occurs) for various values of $a_{BX}$. Here we show the results for $\zeta=0$ with $m_{B}:m_{X}=23: 40$ {(black solid line)}, $1:1$ {(blue dotted line)} and $87:40$ {(red dashed line)}.}
\label{resonance2} 
\end{figure}

\begin{figure}[t]
\includegraphics[width=7cm]{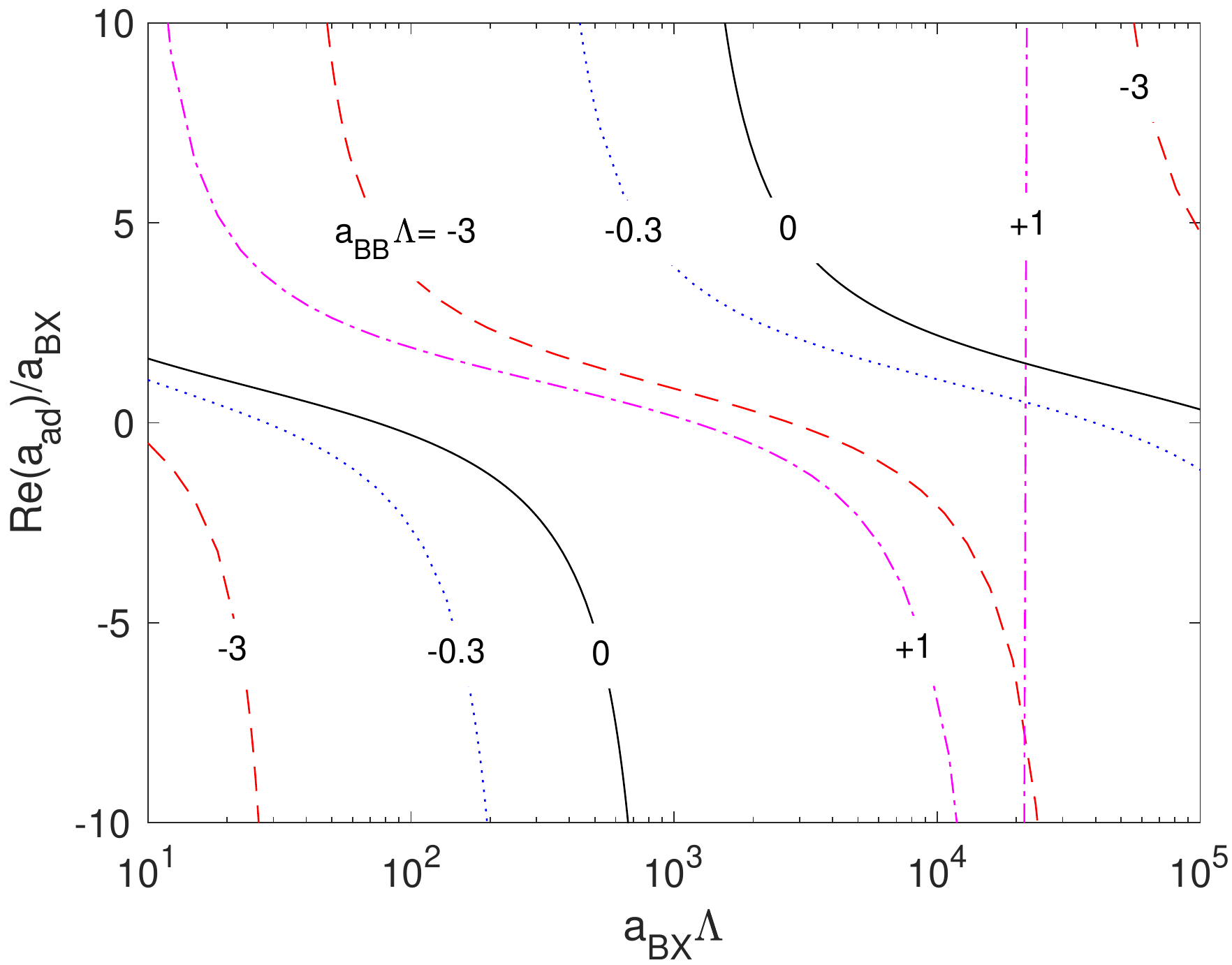} 
\caption{(color online) Ratio $a_{{\rm ad}}/a_{BX}$ as a function of $a_{BX}\Lambda$ for instances $\zeta=0$, $m_{B}=m_{X}$, and $a_{BB}\Lambda=-3$ (red dashed line), -0.3 (blue dotted line), 0 (black solid line), 1 ((magenta dashed-dotted line).}
\label{aadabx} 
\end{figure}

Now we consider systems with non-zero $a_{BB}$. In Fig.~\ref{aadabx}, we plot ${\rm Re}[a_{{\rm ad}}]$ as a function of $a_{BX}$ with fixed $a_{BB}$. Here we show the results for $m_{B}=m_{X}$ as a instance. When $a_{BB}$ is finite, the atom--dimer scattering resonance clearly still occurs. Nevertheless, the variance of $a_{BB}$ in the small region with $|a_{BB}|\lesssim1$ shifts the resonance points by several orders of magnitudes. For instance, when $a_{BB}$ changes from $-3/\Lambda$ to $0$ and then to $+1/\Lambda$, one resonance point is shifted from a point where $a_{BX}<10^{2}/\Lambda$ to $a_{BX}\approx10^{3}/\Lambda$ and then to {$a_{BX}\approx10^{4}/\Lambda$}.

\begin{figure}
\centering \includegraphics[width=7cm]{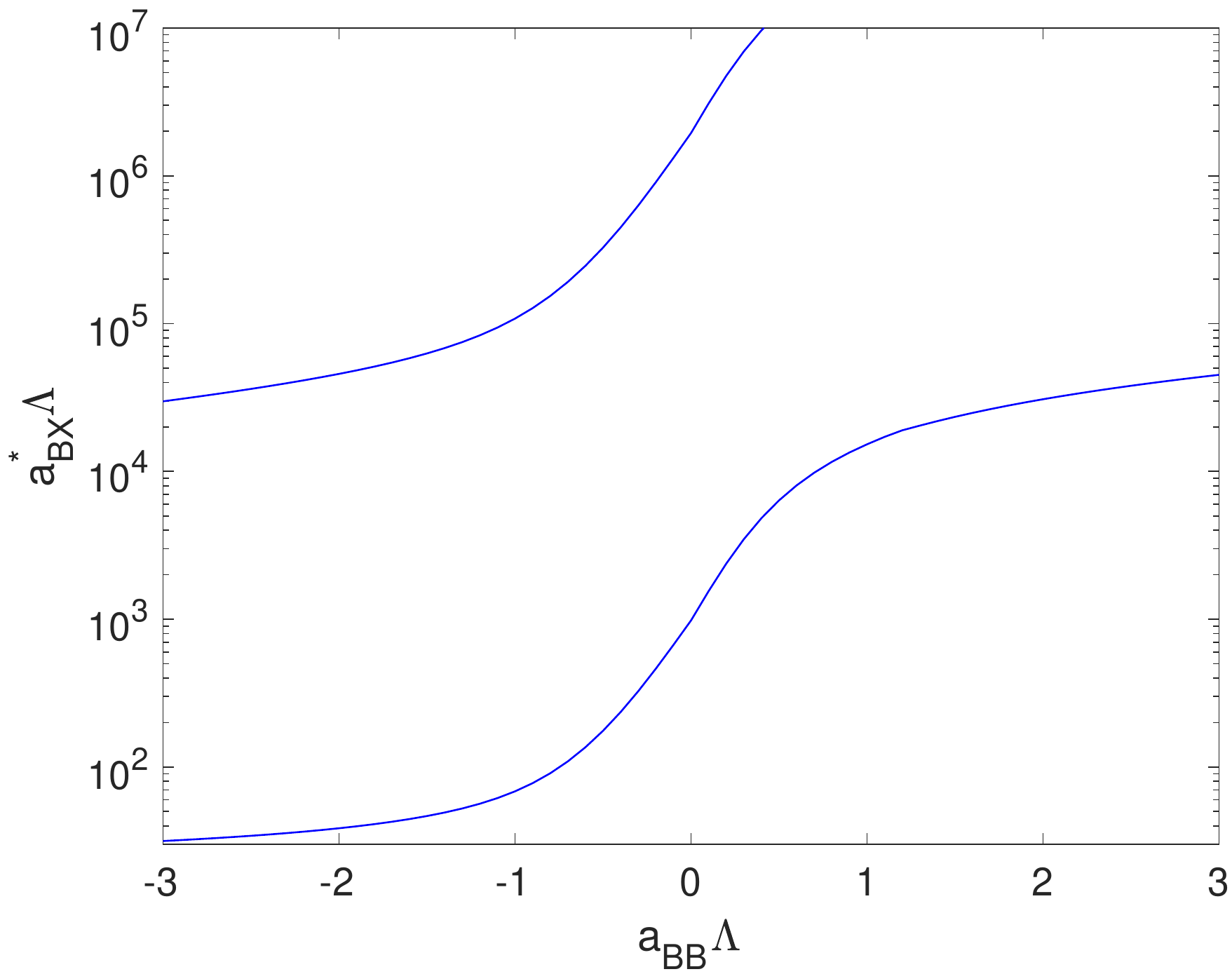}
\caption{Resonance position of the ${\rm Re}[a_{{\rm ad}}]$-$a_{BX}$ function (i.e., values of $a_{BX}$ for which resonance occurs) for various values of $a_{BX}$. Here we show the results for $\zeta=0$ and $m_{B}=m_{X}$.}
\label{resonance3} 
\end{figure}

To further confirm this shift in atom--dimer resonance point, in Fig.~\ref{resonance3}, we plot the resonance position of the ${\rm Re}[a_{{\rm ad}}]$-$a_{BX}$ curve (i.e., the value of $a_{BX}$ for which resonance occurs) for instances with $m_{B}=m_{X}$ and $a_{BB}\in[-3/\Lambda,+3/\Lambda]$. Indeed, this figure may be obtained by exchanging the vertical and perpendicular coordinates of Fig.~\ref{resonance2}. In Fig.~\ref{resonance3}, when $a_{BB}$ varies in the small region, the resonance points of the ${\rm Re}[a_{{\rm ad}}]$-$a_{BX}$ function always shifts by about two orders of magnitudes. For $m_{B}:m_{X}=23:40$ and $87:40$, we have quite similar results.

The above results show that, due to the ``$a_{BB}$-induced\char`\"{} resonance effect discussed in the above subsection, the ${\rm Re}[a_{{\rm ad}}]$-$a_{BX}$ function has a sensitive dependence on the boson--boson scattering length $a_{BB}$. To evaluate just the order of magnitude of the resonance point of this function, one still requires to take into account the precise value of $a_{BB}$.

\section{Summary and Discussion}

We calculated the scattering length $a_{{\rm ad}}$ between an ultracold bosonic atom and a shallow dimer formed by an identical bosonic atom and another distinguishable atom, and investigated the influence of the weak intra-species interaction between the two bosonic atoms on the atom--dimer scattering. We found that $a_{{\rm ad}}$ resonantly changes with the boson--boson scattering length $a_{BB}$, even in the region where $a_{BB}$ is as small as the van der Waals length. As a result, a slight variation in $a_{BB}$ modifies the sign and the absolute value of $a_{{\rm ad}}$, as well as the relationship between $a_{{\rm ad}}$ and the large inter-species scattering length $a_{BX}$. Our results show that although the intra-species interaction is weak, it still needs to be taken into account in the relevant quantitative theories for this ultra-cold heteronuclear mixture system. 

Note that our qualitative result is also applicable to systems with three distinguishable atoms 1,2,3, where masses $m_i$ of atoms $i\ (i=1,2,3)$ and scattering lengths $a_{ij}$ $(i,j=1,2,3)$ between atoms $i$ and $j$ satisfy $m_1=m_2$, $a_{23}=a_{31}\gg r_{\rm vdW}$, and $|a_{12}|\lesssim r_{\rm vdW}$. 
For this system, as discussed in Appendix A, there can be both elastic and inelastic scattering between atom 1 and the shallow dimer formed by 2 and 3. After inelastic scattering, atoms 1 and 3 form a shallow dimer and atom 2 becomes free. Furthermore, the scattering lengths $a_{\rm ad}^{\rm (e)}$ and $a_{\rm ad}^{\rm (i)}$ for elastic and inelastic scattering satisfy $a_{\rm ad}^{\rm (e,i)}=(a_{\rm ad}^{(B)}\pm a_{\rm ad}^{(F)})/2$, with $a_{\rm ad}^{(B)}$ $(a_{\rm ad}^{(F)})$ being the atom--dimer scattering length for instances when 1 and 2 are identical bosons (fermions). Specifically, $a_{\rm ad}^{(B)}$ is the scattering length calculated in the above sections. As shown above, $a_{\rm ad}^{(B)}$ resonantly changes with $a_{12}$, even if $|a_{12}|$ is very small. Alternatively, it was shown that $a_{\rm ad}^{(F)}=\xi a_{23}$, with $\xi$ being a finite constant determined by $m_1/m_3$ \cite{Petrov2003}. Therefore, our calculation in the foregoing sections implies that a weak interaction between atoms 1 and 2 induces a significant effect for both elastic and inelastic scattering lengths $a_{\rm ad}^{\rm (e)}$ and $a_{\rm ad}^{\rm (i)}$. 

\begin{acknowledgments}
We thank Shizhong Zhang, Hui Zhai, and Ren Zhang for helpful discussions and important suggestions. This work has been supported by the Natural Science Foundation of China under Grant Nos. 11434011, 11674393, and 11604300, and by NKBRSF of China under Grant No. 2012CB922104, the Fundamental Research Funds for the Central Universities, and the Research Funds of Renmin University of China under Grant No. 16XNLQ03. 
\end{acknowledgments}


\appendix

\begin{widetext}

\section{STM-Equations}

\label{STM} 
In this Appendix, we derive the STM equations for the calculation of the atom--dimer scattering length $a_{{\rm ad}}$, i.e., Eqs.~(\ref{stm1}, \ref{stm2}) and Eq.~(\ref{sl}) in the main text. Here we use the approach in our previous work, given in the appendix of Ref.~\cite{Zhang2014}. We label the two bosonic atoms as 1 and 2, and the distinguishable atom $X$ as 3, and model the binary interaction $V_{{ij}}$ ($i,j=1,2,3$) using the Huang--Yang pseudo potential 
\begin{equation}
V=\frac{2\pi a_{{ij}}}{\mu_{{ij}}}\delta\left({\bf r}_{ij}\right)\frac{\partial}{\partial|{\bf r}_{ij}|}\left(|{\bf r}_{ij}|\cdot\right),
\end{equation}
where ${\bf r}_{ij}$ ($i,j=1,2,3$) is the relative position of the atoms $i$ and $j$ with $a_{12}=a_{{BB}}$ and $a_{23}=a_{31}=a_{{BX}}$, and $\mu_{ij}=m_{i}m_{j}/(m_{i}+m_{j})$ is the reduced mass of atoms $i$ and $j$, with $m_{j}$ being the mass of atom $j$. Here we use natural units with $\hbar=m_{B}=1$, and thus we have $m_{1}=m_{2}=1$ and $m_{3}=M\equiv m_{X}/m_{B}$.

In the following, we first ignore the Bose statistics of atoms 1 and 2, and calculate the atom--dimer scattering length for the three distinguishable particles. We then take into account the identity of the two bosonic atoms and derive $a_{{\rm ad}}$ for the system studied in the main text.

\subsection{Three distinguishable atoms}

We first assume all three atoms to be distinguishable particles, and study the threshold scattering between atom 1 and the shallow dimer formed by atoms 2 and 3. Since we have $a_{23}=a_{31}=a_{{BX}}$, in this scattering process there are two possible output states, which are degenerate, i.e., the state where atom 1 is free while 2 and 3 form a shallow dimer, and the state where atom 2 is free while 1 and 3 form a shallow dimer. The amplitude for the collision with these two output states is described by the elastic scattering length $a_{{\rm ad}}^{({\rm e})}$ and inelastic scattering length $a_{{\rm ad}}^{({\rm i})}$, respectively. The corresponding atom--dimer scattering length is defined as ($\hbar=1$) 
\begin{eqnarray}
a_{{\rm ad}}^{({\rm e})} & = & 4\pi^{2}m_{\text{ad}}\langle\Psi_{0}^{(1)}|V^{(1)}|\Psi\rangle;\nonumber \\
a_{{\rm ad}}^{({\rm i})} & = & 4\pi^{2}m_{\text{ad}}\langle\Psi_{0}^{(2)}|V^{(2)}|\Psi\rangle,\label{aaddis}
\end{eqnarray}
where $m_{{\rm ad}}=m_{1}(m_{2}+m_{3})/(m_{1}+m_{2}+m_{3})$ is the atom--dimer reduced mass, $V^{(1)}=V_{{12}}+V_{{31}}$ and $V^{(2)}=V_{{23}}+V_{{12}}$. Here $|\Psi_{0}^{(1)}\rangle$ and $|\Psi_{0}^{(2)}\rangle$ are the two output states given by
\begin{eqnarray}
|\Psi_{0}^{(1)}\rangle & = & |{\bf K}=0\rangle_{1-23}|\phi_{b}\rangle_{23};\label{psi01}\\
|\Psi_{0}^{(2)}\rangle & = & |{\bf K}=0\rangle_{2-31}|\phi_{b}\rangle_{31},\label{psi02}
\end{eqnarray}
with $|{\bf K}\rangle_{i-jk}$ being the eigen-state of the relative momentum of atom $i$ and the center-of-mass of atoms $j$ and $k$, and $|\phi_{b}\rangle_{jk}$ the shallow bound states of atoms $j$ and $k$. Note that $|\Psi_{0}^{(1)}\rangle$ is also the incident state of our scattering process. In Eq.~(\ref{aaddis}), $|\Psi\rangle$ is the atom--dimer scattering state corresponding to the incident state $|\Psi_{0}^{(1)}\rangle$, and is given by \cite{fewbodybook}
\begin{equation}
|\Psi\rangle=\lim_{\varepsilon\rightarrow0^{+}}\frac{i\varepsilon}{E+i\varepsilon-H}|\Psi_{0}^{(1)}\rangle\equiv\lim_{\varepsilon\rightarrow0^{+}}|\Psi(\varepsilon)\rangle,\label{psi}
\end{equation}
where $E=-1/(2\mu_{23}a_{23}^{2})$ is the scattering energy, and $H=T+V_{12}+V_{13}+V_{23}$ the total Hamiltonian for our three-body problem with $T$ being the total kinetic energy of the relative motion of the three atoms.

Similar as in Ref.~\cite{Zhang2014}, to calculate $a_{{\rm ad}}^{({\rm e})}$ and $a_{{\rm ad}}^{({\rm i})}$, we introduce here three functions
\begin{eqnarray}
\eta^{(1)}({\bf K},\varepsilon) & = & \frac{2\pi a_{23}}{\mu_{23}}\left[\partial_{|{\bf r}|}|{\bf r}|\cdot{}_{23}\!\langle{\bf r}|{}_{1-23}\!\langle{\bf K}|\Psi(\varepsilon)\rangle\right]_{{\bf r}=0};\label{eta1}\\
\eta^{(2)}({\bf K},\varepsilon) & = & \frac{2\pi a_{31}}{\mu_{31}}\left[\partial_{|{\bf r}|}|{\bf r}|\cdot{}_{31}\!\langle{\bf r}|{}_{2-31}\!\langle{\bf K}|\Psi(\varepsilon)\rangle\right]_{{\bf r}=0};\label{eta2}\\
\eta^{(3)}({\bf K},\varepsilon) & = & \frac{2\pi a_{12}}{\mu_{12}}\left[\partial_{|{\bf r}|}|{\bf r}|\cdot{}_{12}\!\langle{\bf r}|{}_{3-12}\!\langle{\bf K}|\Psi(\varepsilon)\rangle\right]_{{\bf r}=0},\label{eta3}
\end{eqnarray}
where $|{\bf r}\rangle_{ij}$ ($i,j=1,2,3$) is the eigen-state of the relative position of atoms $i$ and $j$. Clearly, we have 
\begin{equation}
_{ij}\!\langle{\bf r}|{}_{k-ij}\!\langle{\bf K}|V_{ij}|\Psi(\varepsilon)\rangle=\delta({\bf r})\eta^{(k)}({\bf K},\varepsilon),\label{etab}
\end{equation}
with $(i,j,k)=(1,2,3)$, $(2,3,1)$ or $(3,1,2)$. We can further define two auxiliary functions $A^{{\rm (e)}}({\bf K},\varepsilon)$ and $A^{{\rm (i)}}({\bf K},\varepsilon)$ via 
\begin{eqnarray}
\eta^{(1)}({\bf K},\varepsilon) & = & -\frac{\sqrt{2\pi}}{\mu_{23}\sqrt{a_{23}}}\left[\delta({\bf K})+\frac{A^{({\rm e})}({\bf K},\varepsilon)}{2\pi^{2}\left(2i\varepsilon\mu_{1-23}-|{\bf K}|^{2}\right)}\right];\label{bigae}\\
\eta^{(2)}({\bf K},\varepsilon) & = & -\frac{A^{({\rm i})}({\bf K},\varepsilon)}{\sqrt{2}\pi^{\frac{3}{2}}\mu_{31}\sqrt{a_{31}}\left(2i\varepsilon\mu_{2-31}-|{\bf K}|^{2}\right)}.\label{bigai}
\end{eqnarray}
Using the approach in the appendix of Ref.~\cite{Zhang2014}, we can directly prove that $A^{({\rm e})}({\bf K},\varepsilon)$ and $A^{({\rm i})}({\bf K},\varepsilon)$ are related to $a_{{\rm ad}}^{({\rm e})}$ and $a_{{\rm ad}}^{({\rm i})}$ via 
\begin{eqnarray}
a_{{\rm ad}}^{({\rm e})} & = & \lim_{\varepsilon\rightarrow0^{+}}A^{({\rm e})}({\bf K}=0,\varepsilon);\label{aeexp}\\
a_{{\rm ad}}^{({\rm i})} & = & \lim_{\varepsilon\rightarrow0^{+}}A^{({\rm i})}({\bf K}=0,\varepsilon).\label{aiexp}
\end{eqnarray}

Now we derive the equations of the functions $A^{({\rm e})}({\bf K},\varepsilon)$, $A^{({\rm i})}({\bf K},\varepsilon)$ and $\eta^{(3)}({\bf K},\varepsilon)$. We first note that Eq.~(\ref{psi}) leads to the result
\begin{equation}
|\Psi(\varepsilon)\rangle=i\varepsilon G_{0}(\varepsilon)|\Psi_{0}^{(1)}\rangle+G_{0}(\varepsilon)[V_{12}+V_{23}+V_{13}]|\Psi\rangle,\label{lps}
\end{equation}
with $G_{0}(\varepsilon)=1/[E+i\varepsilon-T]$. As shown in the appendix of Ref.~\cite{Zhang2014}, using this result we can directly obtain the equations for $\eta^{(i)}({\bf K},\varepsilon)$ ($i=1,2,3$). Substituting Eqs.~(\ref{aeexp}, \ref{aiexp}) into these equations
and using $a_{12}=a_{{BB}}$, $a_{23}=a_{31}=a_{{BX}}$, $m_{1}=m_{2}=1$, and $m_{3}=M$, we obtain 
\begin{eqnarray}
 &  & -\frac{A^{({\rm e})}({\bf K},\varepsilon)}{\sqrt{2a_{BX}}\mu_{BX}\pi^{\frac{3}{2}}\left(2i\varepsilon m_{\rm ad}-|{\bf K}|^{2}\right)}\nonumber \\
 & = & \frac{a_{BX}}{4\mu_{BX}\pi^{2}}\int d\tilde{{\bf K}}\frac{\eta^{(3)}(\tilde{{\bf K}},\varepsilon)}{\left[E+i\varepsilon-\tilde{M}|\tilde{{\bf K}}|^{2}-|{\bf K}+\frac{1}{2}\tilde{{\bf K}}|^{2}\right]}-\frac{\sqrt{a_{BX}}}{\sqrt{\mu_{BX}}\pi^{\frac{3}{2}}}\sqrt{\frac{|{\bf K}|^{2}}{2m_{\rm ad}}-E-i\varepsilon}\frac{A^{({\rm e})}({\bf K},\varepsilon)}{\left[2i\varepsilon m_{\rm ad}-|{\bf K}|^{2}\right]}\nonumber \\
 &  & -\frac{\sqrt{a_{BX}}}{4\sqrt{2}\pi^{\frac{7}{2}}\mu_{BX}^{2}}\int d\tilde{{\bf K}}\frac{A^{({\rm i})}({\bf \tilde{{\bf K}}},\varepsilon)}{\left[E+i\varepsilon-\frac{1}{2m_{\rm ad}}|\tilde{{\bf K}}|^{2}-\frac{1}{2\mu_{BX}}|{\bf K}+\frac{1}{M+1}\tilde{{\bf K}}|^{2}\right]\left[2i\varepsilon m_{\rm ad}-|\tilde{{\bf K}}|^{2}\right]}\nonumber \\
 &  & +i\varepsilon\sqrt{\frac{1}{2\pi^{3}a_{BX}}}\frac{1}{[E\mu_{BX}+i\varepsilon\mu_{BX}-|{\bf K}|^{2}/2](|{\bf K}|^{2}+\frac{1}{a_{BX}^{2}})};\label{stm11}
\end{eqnarray}
\begin{eqnarray}
 &  & -\frac{A^{({\rm i})}({\bf K},\varepsilon)}{\sqrt{2a_{BX}}\mu_{BX}\pi^{\frac{3}{2}}\left(2i\varepsilon m_{\rm ad}-|{\bf K}|^{2}\right)}\nonumber \\
 & = & \frac{a_{BX}}{4\mu_{BX}\pi^{2}}\int d\tilde{{\bf K}}\frac{\eta^{(3)}(\tilde{{\bf K}},\varepsilon)}{\left[E+i\varepsilon-\tilde{M}|\tilde{{\bf K}}|^{2}-|{\bf K}+\frac{1}{2}\tilde{{\bf K}}|^{2}\right]}-\frac{\sqrt{a_{BX}}}{\sqrt{\mu_{BX}}\pi^{\frac{3}{2}}}\sqrt{\frac{|{\bf K}|^{2}}{2m_{\rm ad}}-E-i\varepsilon}\frac{A^{({\rm i})}({\bf K},\varepsilon)}{\left[2i\varepsilon m_{\rm ad}-|{\bf K}|^{2}\right]}\nonumber \\
 &  & -\frac{\sqrt{a_{BX}}}{4\sqrt{2}\pi^{\frac{7}{2}}\mu_{BX}^{2}}\int d\tilde{{\bf K}}\frac{A^{({\rm e})}(\tilde{{\bf K}},\varepsilon)}{\left[E+i\varepsilon-\frac{1}{2m_{\rm ad}}|\tilde{{\bf K}}|^{2}-\frac{1}{2\mu_{BX}}|{\bf K}+\frac{1}{M+1}\tilde{{\bf K}}|^{2}\right]\left[2i\varepsilon m_{\rm ad}-|\tilde{{\bf K}}|^{2}\right]}\nonumber \\
 &  & -\frac{\sqrt{a_{BX}}}{2\sqrt{2}\mu_{BX}^{2}\pi^{\frac{3}{2}}\left[E+i\varepsilon-\frac{1}{4\mu_{BX}}|{\bf K}|^{2}\right]}+i\varepsilon\sqrt{\frac{1}{2\pi^{3}a_{BX}}}\frac{1}{[E\mu_{BX}+i\varepsilon\mu_{BX}-|{\bf K}|^{2}/2](|{\bf K}|^{2}+\frac{1}{a_{BX}^{2}})};\label{stm22}
\end{eqnarray}
and 
\begin{eqnarray}
\eta^{(3)}({\bf K},\varepsilon) 
& = & -\frac{a_{BB}}{2\sqrt{2}\pi^{\frac{7}{2}}\sqrt{a_{BX}}\mu_{BX}}\int d\tilde{{\bf K}}
\frac{A^{({\rm e})}(\tilde{{\bf K}},\varepsilon)+A^{({\rm i})}(\tilde{{\bf K}},\varepsilon)}
{\left[E+i\varepsilon-\frac{1}{2m_{\rm ad}}|\tilde{{\bf K}}|^{2}
-\frac{1}{2\mu_{BX}}|{\bf K}+\frac{M}{M+1}\tilde{{\bf K}}|^{2}\right]
\left[2i\varepsilon m_{\rm ad}-|\tilde{{\bf K}}|^{2}\right]}\nonumber \\
 &  & -\frac{a_{BB}}{\sqrt{2}\pi^{\frac{3}{2}}\sqrt{a_{BX}}\mu_{BX}
\left[E+i\varepsilon-\frac{1}{2\mu_{BX}}|{\bf K}|^{2}\right]}
+a_{BB}\sqrt{\tilde{M}|\tilde{{\bf K}}|^{2}-E-i\varepsilon}\eta^{(3)}({\bf K},\varepsilon).
\label{stm33}
\end{eqnarray}
with $\tilde{M}=(M+2)/4M$ and $\mu_{BX}= M/(M+1)$. Recall that in our current natural units, we have $m_{\rm ad}=(M+1)/(M+2)$.

\subsection{Two identical bosonic atoms and one extra atom}

Now we consider the system studied in the main text, where atoms 1 and 2 are identical bosons. As a result of the bosonic statistics, the incident state of the atom--dimer threshold scattering length is symmetric $|\Phi\rangle\equiv(|\Psi_{0}^{(1)}\rangle+|\Psi_{0}^{(2)}\rangle)/\sqrt{2}$, with $|\Psi_{0}^{(1,2)}\rangle$, being defined in the above subsection. In addition, in this instance there is only one possible output state, which is also $|\Phi\rangle$. Consequently, the atom--dimer scattering length $a_{{\rm ad}}$ is given by 
\begin{eqnarray}
a_{{\rm ad}}=a_{{\rm ad}}^{({\rm e})}+a_{{\rm ad}}^{({\rm i})},\label{aaddef}
\end{eqnarray}
where $a_{{\rm ad}}^{({\rm e})}$ and $a_{{\rm ad}}^{({\rm i})}$ are the elastic and inelastic scattering lengths when the three atoms are distinguishable, as we have defined above. Thus, we can introduce a function 
\begin{eqnarray}
A({\bf K},\varepsilon)=A^{{\rm (e)}}({\bf K},\varepsilon)+A^{{\rm (i)}}({\bf K},\varepsilon),
\end{eqnarray}
with $A^{{\rm (e,i)}}({\bf K},\varepsilon)$ being defined in Eqs.~(\ref{bigae}) and (\ref{bigai}). Eqs.~(\ref{aeexp}), (\ref{aiexp}), and  (\ref{aaddef}) imply 
\begin{eqnarray}
a_{{\rm ad}} & = & \lim_{\varepsilon\rightarrow0^{+}}A({\bf K}=0,\varepsilon).\label{appf}
\end{eqnarray}

Furthermore, summing Eqs.~(\ref{stm11}) and (\ref{stm22}), we can directly obtain integral equations for $A({\bf K},\varepsilon)$ and $\eta^{(3)}({\bf K},\varepsilon)$. Eq.~(\ref{stm33}) can be directly re-written as another integral equation for these two functions. Finally, similar as in Ref.~\cite{Zhang2014}, we can express $A({\bf K},\varepsilon)$ and $\eta^{(3)}({\bf K},\varepsilon)$ as $A({\bf K},\varepsilon)=\sum_{l,m}A_{l,m}(K,\varepsilon)Y_{l}^{m}(\hat{{\bf K}})$ and $\eta^{(3)}({\bf K},\varepsilon)=\sum_{l,m}\eta_{l,m}^{(3)}(K,\varepsilon) Y_{l}^{m}(\hat{{\bf K}})$, respectively, with $K=|{\bf K}|$ and $Y_{l}^{m}$ being the spherical harmonic function associated with the direction of ${\bf K}$. Substituting these expressions into the two integral equations that we have just obtained, we obtain the equations for the components $A_{l,m}$ and $\eta_{l,m}^{(3)}$. We find that the equations for different values of $l$ and $m$ decouple. In addition, Eq.~(\ref{appf}) may be written $a_{{\rm ad}}=\lim_{\varepsilon\rightarrow0^{+}}A_{0,0}(K=0,\varepsilon)/\sqrt{4\pi}$. Thus, to calculate $a_{{\rm ad}}$, we only need to solve the two integral equations for $A_{0,0}$ and $\eta_{0,0}^{(3)}$. Making a simplification of notation, $A_{0,0}(K,\varepsilon)/ \sqrt{4\pi}\rightarrow A(K,\varepsilon)$ and $\eta_{0,0}^{(3)}(K,\varepsilon)/ \sqrt{4\pi}\rightarrow\eta(K,\varepsilon)$, we find that these two equations are just Eqs.~(\ref{stm1}) and (\ref{stm2}) of the main text, and Eq.~(\ref{appf}) can be written as in Eq.~(\ref{sl}).
\end{widetext}



\begin{thebibliography}{10}
\bibitem{Efimov1970} V. Efimov, Phys. Lett. B \textbf{33}, 563 (1970).

\bibitem{Efimov1971} V. Efimov, Sov. J. Nucl. Phys. \textbf{12},
589 (1971).

\bibitem{Efremov2009} M. A. Efremov, L. Plimak, B. Berg, M. Yu. Ivanov,
and W. P. Schleich, Phys. Rev. A \textbf{80}, 022714 (2009).

\bibitem{HHP2010} K. Helfrich, H.-W. Hammer, and D. S. Petrov, Phys.
Rev. A \textbf{81}, 042715 (2010).

\bibitem{bozhao} J. Rui, H. Yang, L. Liu, D. Zhang, Y. Liu, J. Nan,
Y. Chen, B. Zhao and J. Pan, Nat. Phys. \textbf{13}, 699 (2017).

\bibitem{shizhong} S. Zhang and T. Ho, New. Jour. Phys., \textbf{13}
055003 (2011).

\bibitem{Cui2014} X. Cui, Phys. Rev. A \textbf{90}, 041603 (2014),

\bibitem{Zhang2014} R. Zhang, W. Zhang, H. Zhai, and P. Zhang, Phys.
Rev. A \textbf{90}, 063614 (2014).

\bibitem{Efimov1979} V. Efimov, Sov. J. Nuc. Phys. \textbf{29}, 546
(1979) {[}Yad. Fiz. \textbf{29}, 1058 (1979){]}.

\bibitem{Bedaque1999} P. F. Bedaque, H.-W. Hammer, U. van Kolck,
Nucl.Phys. A \textbf{646}, 444 (1999); Phys. Rev. Lett. \textbf{82},
463 (1999). 

\bibitem{Petrov2004BBB} D. S. Petrov, Phys. Rev. Lett. \textbf{93},
143201 (2004),


\bibitem{Petrov2003} D. Petrov, Phys. Rev. A \textbf{67}, 010703
(2003),

\bibitem{Petrov2005} D. S. Petrov, C. Salomon, and G. V. Shlyapnikov,
Phys. Rev. A \textbf{71}, 012708 (2005),

\bibitem{Levinsen2009} J. Levinsen, T. G. Tiecke, J. T. M. Walraven,
and D. S. Petrov, Phys. Rev. Lett. \textbf{103}, 153202 (2009),

\bibitem{Levinsen2011} J. Levinsen and D. S. Petrov, Eur. Phys. J.
D \textbf{65}, 67 (2011),

\bibitem{Iskin2010} M. Iskin, Phys. Rev. A \textbf{81}, 043634 (2010),

\bibitem{Alzetto2010} F. Alzetto, R. Combescot, X. Leyronas, Phys.
Rev. A \textbf{82}, 062706 (2010),

\bibitem{Alzetto2012} F. Alzetto, R. Combescot, X. Leyronas, Phys.
Rev. A \textbf{86}, 062708 (2012), 

\bibitem{Bour2012} S. Bour, H.-W. Hammer, D. Lee, and Ulf-G. Mei$\beta$ner,
Phys. Rev. C \textbf{86}, 034003 (2012), 

\bibitem{Rb85Li7} R.A.W. Maier, M. Eisele, E. Tiemann, and C. Zimmermann,
Phys. Rev. Lett. \textbf{115}, 043201 (2015).

\bibitem{RbK41} G. Barontini, C. Weber, F. Rabatti, J. Catani, G.
Thalhammer, M. In- guscio, F. Minardi, Phys. Rev. Lett. \textbf{103},
043201 (2009).

\bibitem{RbK40a} J. J. Zirbel, K.-K. Ni, S. Ospelkaus, J. P. DIncao,
C. E. Wieman, J. Ye, and D. S. Jin, Phys. Rev. Lett. \textbf{100},
143201 (2008).

\bibitem{RbK40} R. S. Bloom, M. Hu, Tyler, D. Cumby, and D. S. Jin,
Phys. Rev. Lett. \textbf{111}, 105301 (2013).

\bibitem{RbK40b} M. Hu, R. S. Bloom, D. S. Jin, and J. M. Goldwin,
Phys. Rev. A \textbf{90}, 013619 (2014).

\bibitem{RbSr} V. Barb$\acute{{\rm e}}$, A. Ciamei, B. Pasquiou,
L. Reichs$\ddot{{\rm o}}$llner, F. Schreck, P. S. $\dot{{\rm Z}}$uchowski,
and J M. Hutson, arXiv: 1710.03093.

\bibitem{LiCs0} R. Pires, J. Ulmanis, S. H$\ddot{{\rm a}}$fner,
M. Repp, A. Arias, E. D. Kuhnle, and M. Weidem$\ddot{{\rm u}}$ller,
Phys. Rev. Lett. \textbf{112}, 250404 (2014).

\bibitem{LiCs00} S.-K. Tung, K. Jim$\acute{{\rm e}}$nez-Garc$\acute{{\rm i}}$a,
J. Johansen, C. V. Parker, and C. Chin, Phys. Rev. Lett. \textbf{113},
240402 (2014).

\bibitem{LiCs} J. Ulmanis, S. Hfner, R. Pires, F. Werner, D. S. Petrov,
E. D. Kuhnle, and M. Weidemller, Phys. Rev. A \textbf{93}, 022707
(2016).

\bibitem{LiCsd} S.-K. Tung, K. Jimenez-Garcia, J. Johansen, C. V.
Parker, and C. Chin, Phys. Rev. Lett. \textbf{113}, 240402 (2014).

\bibitem{LiCsb} J. Ulmanis, S. H$\ddot{{\rm a}}$fner, R. Pires,
E. D. Kuhnle, Y. Wang, C. H. Greene, and M. Weidem$\ddot{{\rm u}}$ller,
Phys. Rev. Lett. \textbf{117}, 153201 (2016).

\bibitem{LiCsc} S. H$\ddot{{\rm a}}$fner, J. Ulmanis, E. D. Kuhnle,
Y. Wang, C. H. Greene, and M. Weidem$\ddot{{\rm u}}$ller, Phys. Rev.
A \textbf{95}, 062708 (2017).

\bibitem{LiCse} M. Sun and X. Cui, Phys. Rev. A {\bf 96}, 022707 (2017).


\bibitem{AJP2016} B. Acharya, C. Ji, and L. Platter, Phys. Rev. A
\textbf{94}, 032702 (2016).

\bibitem{STM1957} G. V. Skorniakov and K. A. Ter-Martirosian, Sov.
Phys. JETP \textbf{4}, 648 (1957).

\bibitem{fewbodybook} W. Gl${\ddot{\rm o}}$ckle, {\it The Quantum Mechanical Few-Body Problem},
(Springer, New York, 1983).

\end{thebibliography}
\end{document}